\documentstyle[aps,prd,epsfig]{revtex}

\newdimen\nude\newbox\chek
\def\slash#1{\setbox\chek=\hbox{$#1$}\nude=\wd\chek#1{\kern-\nude/}}

\begin{document}

\title{Thermal photon production rate from non-equilibrium 
  quantum field theory}

\preprint{BI-TP 97/11,  
          LPTHE-Orsay 97-15}

\author{R.~Baier and  M.~Dirks}
\address{
Fakult\"{a}t f\"{u}r Physik, Universit\"{a}t Bielefeld,
D-33501 Bielefeld, Germany}

\author{K.~Redlich}
\address{Institute for Theoretical Physics, University of Wroclaw, \\
PL-50204 Wroclaw, Poland\\and\\
GSI, PF 110552,  D-64220 Darmstadt, Germany}

\author{D.~Schiff}
\address{LPTHE Universit\'e Paris-Sud, B\^atiment 211, F-91405 Orsay, France
  \footnote{Laboratoire associ\'e du Centre National de la Recherche 
  Scientifique}}

\maketitle

\begin{abstract}
 In the framework of  closed time path thermal field theory
 we investigate the production rate of hard thermal photons
 from a QCD plasma away from equilibrium.
 Dynamical screening provides a finite rate for chemically 
 non-equilibrated distributions of quarks and
 gluons just as it does in the equilibrium situation. Pinch singularities
 are shown to be absent in the real photon rate even away from equilibrium. 
\end{abstract}

\section{Introduction}
Ultrarelativistic heavy ion experiments at CERN and RHIC focus on energetic
direct photons as promising observables of the expected QCD plasma phase.
Namely the spectrum of thermal photons could be an observable effect
directly related to the multiparticle dynamics of the plasma system, and
as such has been the subject of major theoretical interest.
As a result the hard thermal photon spectrum has been obtained consistently
in thermal field theory implementing dynamical screening mechanisms in
the framework of hard-thermal-loop (HTL) resummation 
\cite{pisarski:std1,pisarski:std2,pisarski:std3,frenkel:htl1}
under the assumption
that the plasma be in perfect thermal equilibrium
\cite{kapusta:photon,baier:photon}. This later assumption
is to be contrasted with parallel investigations of the plasma expansion
\cite{biro:therma,biro:therma2,wong:therm,geiger:cascch}
from which the picture of a hot gluon gas \cite{shuryak:hotglue.org} 
has emerged by now: quarks
but also gluons are expected to be essentially undersaturated during
important parts of the plasma history. The effect of this undersaturation
on the photon spectrum has been examined in 
\cite{shuryak:hotglue,thoma:nephoton,kaempfer:nephoton,muller:therma} 
on the
basis of folding scaled distributions on conventional matrix elements.
The mass singularity appearing in this process has been cut-off
by hand introducing a thermal quark mass suitably generalized from the
equilibrium one. No attempt has however been made to obtain the necessary 
screening  from a
consistent treatment within thermal field theory so that the
validity of this cut-off prescription seems to be 'unclear' at the
moment. 

In this note we therefore examine in some detail the
generalization of the HTL-resummation program to the case of undersaturated
distribution functions as applied to the hard photon rate.
We do so on the basis of the closed time path approach
for field theoretic systems out-of-equilibrium. We take into account 
the appearance of
recently found pinch singularities \cite{altherr:pinch1,altherr:pinch2} 
which can be shown to provide  extra contributions
to physical rates in out-of-equilibrium situations \cite{baier:pirho}.
We however demonstrate that these can effectively be neglected in
the present case of real photon production.
Our derivation  is also easily applied to the particular case of equilibrium
distributions with non-vanishing chemical potential
\cite{greiner:muphoton,thoma:muphoton}. 
Here the photon rate can be explicitly 
shown to be free of infrared singularities as obtained in
\cite{thoma:muphoton} only by numerical integration of the phase space. 

After giving a short review of the theoretical framework used 
we start with a close look on the contribution from the soft part
of phase space in section \ref{soft part} and add the contribution
from the hard part in section \ref{hard part}. Finally the result
is obtained  and discussed in section \ref{result}.

\section{Thermal photon rate} \label{sec:calc}

We adopt the real time formulation of quantum field theory
\cite{landsmann:rev,lebellac:book} in order
to calculate the photon production rate. 
The Keldysh variant is
appropriate for a description of systems away from
equilibrium \cite{chou:noneq}. Both propagators and self-energies
acquire a $2\times 2$ matrix structure in this formalism. 
The components 
$(12)$  and $(21)$ of the  self-energy matrix are related to the 
emission and absorption probability
of the particle species under consideration 
\cite{chou:noneq,heinz:transport1,calzetta:noneq1}. In the
present context we  treat the photons as external probes, i.e., thermally
decoupled from the quark-gluon plasma. 
To lowest order they are produced from annihilation and Compton
processes
\begin{equation}
 q + \bar q \to g + \gamma, \qquad
 q (\bar q) + g \to q (\bar q) + \gamma,
\end{equation}
but there are essentially no back-reactions that would absorb photons
present in the medium. 
The rate of photon emission can thus be calculated as 
\begin{equation}\label{eq:rate}
 E\frac{dR}{d^3 q} = \frac{i}{2(2\pi)^3} {\Pi_{12}}_\mu^\mu (q),
\end{equation}
from the trace of the (12)-element $\Pi_{12}$ of the photon-polarization 
tensor. 

\begin{figure}
\centering
\epsfig{file=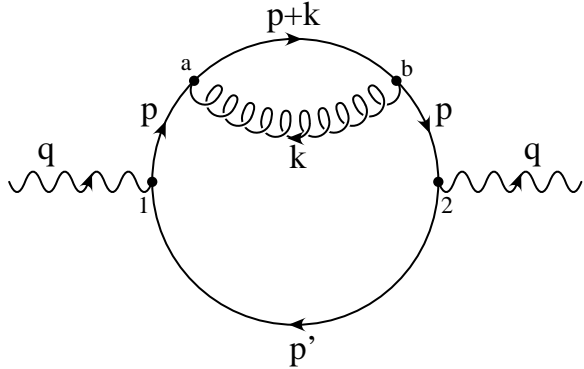,width=8cm}
\caption{\label{fig:fix} The photon polarization tensor $\Pi_{12}$ 
 to first order in  $\alpha_s$ for
 real photon production. The momentum labels define the notation used
 in the text. External vertices are to be of type 1 and 2, respectively,
 internal vertices are to be summed over $a,b = 1,2$. }
\end{figure}

We calculate $\Pi_{12}$ to leading order in the strong
interaction of quarks and gluons. The relevant diagram in fixed order
perturbation theory is shown in Fig.\ref{fig:fix}. The relevant (12)
and (21) propagator components depend
on the (off-equilibrium) distributions of quarks and gluons.  

Having in mind locally thermalized but undersaturated distributions
for quarks and gluons we neglect higher than 
second order terms in the cumulant expansion for the off-equilibrium 
correlations \cite{chou:noneq}. Separating a macroscopic scale
$X$ from a fast microscopic scale $p$ in the sense of a Wigner transform
and applying a gradient expansion to the former, the relevant propagators 
can be shown to depend on a Wigner distribution $n(X,p)$ formerly in the
same way as in equilibrium \cite{lebellac:pinch}. 
For the bosonic case
all relevant quantities have been listed in \cite{baier:pirho}.
Sticking to the conventions used there we  use the 
gluon propagator in Feynman gauge as
\begin{equation}
 iD^{12}_{\mu\nu} (X,p) = - g_{\mu\nu} 2\pi \varepsilon(p_0)\delta(p^2) 
  n(X,p_0),
 \qquad
 iD_{\mu\nu}^{21}(X,p) = iD_{\nu\mu}^{12}(X,-p).
\end{equation}
We choose
\cite{biro:therma,biro:therma2,thoma:nephoton}
\begin{equation} \label{eq:bdist}
 n(X,p_0) = n(p_0) = \left\{
 \begin{array}{cc}
  \lambda_g n_B(|p_0|) & p_0 > 0\\
  -(1+ \lambda_g n_B(|p_0|)) & p_0 < 0
 \end{array} \right. ,
\end{equation}
with $n_B(|p_0|)$ the Bose distribution in equilibrium and fugacities
$\lambda_g$ parameterizing the deviation from chemical equilibrium for gluons.  
The dependence on
the macroscopic variable $X$ is chosen to reside in the
parameters $\lambda_g=\lambda_g(X)$ and $T=T(X)$. For our approach
to be valid we assume that this macroscopic scale
is large even against the soft scale $(gT)^{-1}$ for a small 
QCD coupling $g$ so that we differ in this respect from the assumptions
discussed in \cite{blaizot:noneq1}.

Fermionic quantities are obtained by substituting  
fermion distributions $\tilde n(X,p_0) = \tilde n(p_0)$
obtained from (\ref{eq:bdist}) through the replacement 
$ \lambda_g n_B \to - \lambda_q n_F$,
with $n_F(p_0)$ the Fermi-Dirac distribution and $\lambda_q$ 
parameterizing accordingly the deviation from equilibrium for the 
quark- as well as antiquark-distributions:
\begin{equation} \label{eq:bdistf}
 \tilde n(X,p_0) = \tilde n(p_0) = \left\{
 \begin{array}{cc}
  \lambda_q n_F(|p_0|) & p_0 > 0\\
  1 -  \lambda_q n_F(|p_0|)) & p_0 < 0
 \end{array} \right. .
\end{equation}
We make use of the (12)-part of the fermion propagator for massless quarks
which reads
\begin{equation}
 iS_{12}(p) = - \slash p 2\pi \delta(p^2) \varepsilon(p_0) \tilde n(X,p_0),
 \qquad
 iS_{21}(p) = -iS_{12} (-p).
\end{equation}

All quantitative results will be given for the
choice of distributions Eqs.(\ref{eq:bdist}) and (\ref{eq:bdistf}). They thus 
fit into the framework of the model developed in 
\cite{biro:therma,biro:therma2} and unnecessary technical
complications are avoided because of the factorized dependence
on fugacities. 
Our main result, i.e. the screening of the quark mass singularity, is
however rather independent of this choice. It applies in particular 
equally well to the case of J\"uttner distributions 
\begin{equation}\label{eq:juettner}
 n_J(|p_0|) = \frac{\lambda}{e^{|p_0|/T} \pm \lambda} ,
\end{equation}
used e.g. in the analysis by Strickland \cite{strickland:nephoton}. 
Here the functional dependence on the fugacity models
the introduction of a chemical potential. 
However this parameter takes the same value for fermions and
anti-fermions, $\lambda=\lambda_q$, and it is also attributed to the
gluons, $\lambda=\lambda_g$. This is to be contrasted with the equilibrium
distributions, where the chemical potential $\mu$ enters with different
signs for particles and antiparticles, and vanishes for gluons.

\begin{figure}
\centering
\epsfig{file=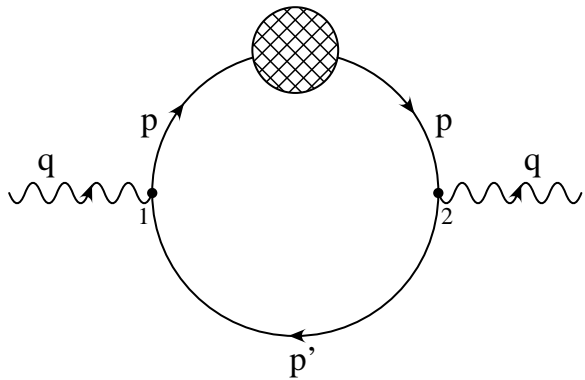,width=8cm}
\caption{\label{fig:res} The photon polarization tensor $\Pi_{12}$ 
 for soft momentum $p\sim gT$. 
 The blob indicates  the HTL-resummed quark propagator.}
\end{figure}

For real photons the self-energy $\Pi_{12}$ in (\ref{eq:rate}) receives 
contributions only from spacelike loop-momenta $p^2  \le 0$. 
The fixed order result from the
diagram Fig.\ref{fig:fix} turns out to contain IR singular contributions 
from small $p^2 \to 0$. For the equilibrium case it has been shown that
this unphysical behavior can be cured taking HTL-contributions from all
orders of the quark self-energy consistently into account
\cite{kapusta:photon,baier:photon}. 
In the spirit of the HTL-resummation program it is important to distinguish
regions of hard from those of soft momenta \cite{braaten:axion}. 
We choose to separate
these scales along the line $p^2 = - k_c^2$ with $k_c^2$ to be chosen on 
an intermediate scale \cite{kapusta:photon}:
\begin{equation}
  -p^2_{hard} \sim T^2 \ge  k_c^2 \sim gT^2 \ge -p^2_{soft} \sim g^2 T^2.  
\end{equation}
For soft momenta resummation of leading HTL-contributions from 
one-loop self-energy insertions is accomplished by substituting the effective
resummed propagator as indicated by a  blob in Fig.\ref{fig:res}.
The resulting partial rate is IR finite and is shown to exactly match the
hard contribution thereby introducing the thermal mass-parameter $m_q$ as an 
effective IR cut-off. 

Turning to the off-equilibrium situation now our first task is to 
determine the appropriate generalization of the effective
resummed propagator to be used on the soft scale.

\subsection{soft part} \label{soft part}

In the non-equilibrium case the analogous resummation program is complicated
by the appearance of extra terms already at fixed one-loop order. Being 
proportional to products of basic retarded  and advanced propagators
\begin{equation}\label{eq:pinch}
   \Delta_R(p)\Delta_A(p) =  \frac{1}{p^2 + i\varepsilon \varepsilon(p_0)}
    \frac{1}{p^2 -  i\varepsilon \varepsilon(p_0)} = \frac{1}{(p^2)^2 + 
  \varepsilon^2}, 
\end{equation}
these terms are in addition plugged with pinch singularities for 
$p^2 = 0$ \cite{altherr:pinch1}.
The resummed propagator for the off-equilibrium situation
was discussed in \cite{altherr:pinch2} for the scalar case with
the objective of providing a self consistent cut-off for these pinch
contributions.
Generalizing to the
fermionic case resumming successive (one-loop) self-energy insertions
\begin{equation}\label{eq:sres}
 iS_{12}^\star (p) = iS_{12}^0 + iS_{1a}^0(-i\Sigma_{ab}) iS^0_{b2} + 
  iS^0_{1a}(-i\Sigma_{ab}) iS^0_{bd} (-i\Sigma_{de}) iS^0_{e2} + \cdots
\end{equation}
may be rearranged into the effective propagator
\begin{eqnarray}\label{eq:deltas.star}
 iS_{12}^\star (p)    &=& - \tilde n(p_0)
 (\frac{i}{\slash p -\Sigma + i\varepsilon p_0} +
        \frac{-i}{\slash p -\Sigma^\star - i\varepsilon p_0})  \nonumber \\
 &+&  \frac{i}{\slash p - \Sigma + i\varepsilon p_0 }
  \left[ (1-\tilde n(p_0)) (-i\Sigma_{12}) + \tilde n(p_0) (-i\Sigma_{21}) \right]
  \frac{-i}  {\slash p - \Sigma^\star -i\varepsilon p_0},
\end{eqnarray}
making use of the identities 
\begin{equation}\label{eq:sig.rel}
\Sigma_{11} = - \Sigma_{22}^\star,  \qquad 
Im \Sigma = - \frac{i}{2} (\Sigma_{12} - \Sigma_{21}), \qquad 
Re \Sigma = Re \Sigma_{11},
\end{equation}
where $\Sigma^\star$ denotes complex conjugation. 

In the expression Eq.(\ref{eq:deltas.star}) $\pm i\varepsilon p_0$ 
is stated explicitly in order to keep track of the retarded or
advanced nature of individual terms for later use, even though the
complex thermal self-energy $\Sigma$ renders this infinitesimal 
displacement superfluous. 
With the help of Eq.(\ref{eq:sig.rel}) $\Sigma$ 
can be decomposed as 
\begin{equation}\label{eq:sigma}
 \Sigma = Re \Sigma_{11}  +  \frac{1}{2} (\Sigma_{12} - \Sigma_{21}).
\end{equation}

Evaluating $\Sigma$
using equilibrium field theory and restricting to the leading HTL-contribution
the first term in (\ref{eq:deltas.star}) reduces to the HTL-resummed
effective propagator as used in the equilibrium calculations
\cite{kapusta:photon,baier:photon}. The second extra term associated
with pinch singularities vanishes in the
equilibrium calculation because of the detailed balance relation 
$\Sigma_{12} =- e^{-p_0/T} \Sigma_{21}$.
It does so as well, for the case of an equilibrium chemical
potential $\mu$ applied to the quark distribution.
In the general non-equilibrium situation a non-vanishing self-energy $\Sigma$
provides a self consistent cut-off for the pinch singularity in 
Eq.(\ref{eq:deltas.star}), as it was already 
observed by Altherr \cite{altherr:pinch2}.

We proceed by evaluating the different terms of Eq.(\ref{eq:deltas.star})
in the spirit of the HTL-approximation, i.e., assuming the external
momentum of the propagator to be soft and focusing only on leading 
self-energy contributions
resulting from hard loop momenta.
First, the imaginary part of $\Sigma$, Eq.(\ref{eq:sigma}), is linear 
in the distributions \cite{lebellac:pinch} 
\begin{eqnarray}
 - i\Sigma^- (p) &\equiv& -\frac{i}{2} (\Sigma_{12}(p) - \Sigma_{21}(p) ) \nonumber\\
 \label{eq:sigdiff}
  &=&   - g^2 C_F \int\frac{d^4 k}{(2\pi)^2} (\slash p + \slash k) \delta(k^2)
 \varepsilon(k_0) \delta( (p+k)^2) \varepsilon(p_0+k_0)
 [ n(k_0) + \tilde n(p_0+k_0)],
\end{eqnarray}
with a color-factor $C_F=4/3$ for $N_c=3$ colors. 
Focusing on the leading contribution only, dimensional and angle integrations
in Eq.(\ref{eq:sigdiff}) decouple,
\begin{equation}\label{eq:imhtl}
 \left. -i\Sigma^- (p) \right|_{HTL} = -\frac{g^2 C_F}{2\pi} \int_0^\infty 
  dE_k E_k (n(E_k) + 
  \tilde n(E_k)) \int \frac{d\Omega}{4\pi} \slash {\hat K}~ \delta(P\cdot\hat K) , 
\end{equation}
with $\hat K = (1,\hat k)$  a light-like vector with spatial direction of
$\vec  k$.
In a similar way also the real part of $\Sigma$ in Eq.(\ref{eq:sigma})
can be analyzed in the HTL-approximation. It is found to differ from
the imaginary part Eq.(\ref{eq:imhtl}) only in the angular integrand.

The only difference to the equilibrium situation is thus seen to reside in the
modified distributions to be integrated over $E_k=|\vec k|$ in 
Eq.(\ref{eq:imhtl}). 
The combination of these functions present in Eq.(\ref{eq:imhtl}) is
important for the following discussion. Here it defines the 
thermal mass parameter $m_q$ generalized from the equilibrium case as:
\begin{equation}\label{eq:mfne}
 m_q^2 = \frac{g^2}{2\pi^2} C_F \int_0^\infty E dE (n(E) + \tilde n(E))
  = \frac{g^2 T^2}{12} C_F \left(\lambda_g + \frac{\lambda_q}{2}\right),
\end{equation}
where in the second step we perform the integration for the
specific choice of distributions, Eqs.(\ref{eq:bdist}) and (\ref{eq:bdistf}),
respectively.
There is a factor of 2 in Eq.(\ref{eq:mfne}) resulting from equal 
contributions for positive as well as negative $k_0$ to the integral in 
Eq.(\ref{eq:sigdiff}). In case of an equilibrium chemical potential 
$2(n(E)+\tilde n(E))$ is replaced by  
$2n_B(E) + n_F(E-\mu) + n_F(E+\mu)$ in the integrand of Eq.(\ref{eq:mfne}).

We now turn to evaluate the second, additional contribution to the propagator
in Eq.(\ref{eq:deltas.star}) in HTL-approximation. 
The coefficient in square brackets can be decomposed as
\begin{equation}\label{eq:pinchdec}
 (1-\tilde n(p_0)) \Sigma_{12} + \tilde n(p_0) \Sigma_{21} = 
  (1-2\tilde n(p_0)) \Sigma^- + \Sigma^+,
\end{equation}
where $\Sigma^-$ has already been evaluated in Eq.(\ref{eq:imhtl})
and 
\begin{equation}
 \Sigma^+ = \frac{1}{2}(\Sigma_{12} + \Sigma_{21})
\end{equation}
is demonstrated in the following to be non-leading.
It contains non-linear terms in the distribution functions:
\begin{eqnarray}
 -i\Sigma^+ &=&  g^2 C_F \int\frac{d^4 k}{(2\pi)^2} (\slash p+\slash k) \varepsilon(k_0)
  \delta(k^2) \varepsilon(p_0+k_0) \delta((p+k)^2) \nonumber \\
  & &\qquad 
  [\tilde n(p_0+k_0) - n(k_0) + 2n(k_0)\tilde n(p_0+k_0) ],
\end{eqnarray}
and keeping only leading contributions the angular integral decouples 
as before. The accompanying dimensional integral turns out to be suppressed
by one power of $g$ with respect to the result (\ref{eq:mfne}) for soft
external $p,p_0 \sim gT$ and hard loop momentum $k\sim T$:
\begin{eqnarray}
 & &g^2 \int_0^\infty dk k 
   [\tilde n(p_0+k) - \tilde  n(k-p_0)] (1+2n(k)) \nonumber\\ 
 & &\quad\simeq 2 p_0 g^2 \int_0^\infty dk k  \frac{\partial}{\partial k} 
  \tilde n(k) (1+2n(k))  \sim O(g^3 T^2) .
\end{eqnarray}
This causes the contribution of $\Sigma^+$ in the resummed propagator 
Eq.(\ref{eq:deltas.star}) being of order $ O(1/T)$ to be
dominated by the leading contributions from $\Sigma^-$ and $Re \Sigma_{11}$
which are of order $O\left(1/(gT)\right)$. 
Neglecting therefore $\Sigma^+$ in 
Eq.(\ref{eq:pinchdec}) and substituting the remainder for the square 
brackets in $S^\star_{12}$
Eq.(\ref{eq:deltas.star}) the resummed propagator may be rearranged
into a form similar to the equilibrium case \cite{baier:photon} :
\begin{equation}\label{eq:sstarhtl}
 \left. iS_{12}^\star \right|_{HTL}(p) = \frac{1}{2} \varepsilon(p_0)
 \left[ (\gamma_0 - \vec\gamma \hat p) Im \frac{1}{\tilde D_+} + 
 (\gamma_0 + \vec\gamma\hat p ) Im \frac{1}{\tilde D_-} \right].
\end{equation}
Here $\tilde D_+(p_0,p) , \tilde D_-(p_0,p)$ determining the quark dispersion 
relations
are as in equilibrium \cite{weldon:ferm,weldon:ferm2,pisarski:ferm}, however, 
with the thermal mass now depending
on the off-equilibrium distributions  as given in Eq.(\ref{eq:mfne}). 

It is worth to point out that in equilibrium $iS^\star_{12}(p)$ is proportional
to $n_F(p_0)$, i.e. $\tilde n(p_0)$ in Eq.(\ref{eq:deltas.star}) is 
replaced by $n_F(p_0)$, which in the soft limit $p_0\simeq 0$ is 
approximated by $1/2$. This way the multiplicative factor becomes the same as
in the non-equilibrium case as it is derived in Eq.(\ref{eq:sstarhtl}). 

In order to obtain the partial thermal photon rate  from the
soft part of phase space we come back to Eq.(\ref{eq:rate}) and
evaluate $\Pi_{12}$ from Fig.\ref{fig:res} as 
\begin{equation}
 -i \Pi_{12}(q) = -e^2e_q^2 N_c \int \frac{d^4 p}{(2\pi)^4} Tr \left[ 
 \gamma^\mu \left. iS_{12}^\star(p) \right|_{HTL} \gamma_{\mu} i S_{21}(p-q) + 
 \gamma^\mu iS_{12}(p) \gamma_\mu \left. iS_{21}^\star(p-q) \right|_{HTL} 
 \right],
\end{equation}
with the resummed propagator Eq.(\ref{eq:sstarhtl}) and $e_q$ the quark 
charge.
We now follow analogous steps as in
equilibrium \cite{kapusta:photon,baier:photon}. Restricting  the validity
of the present resummed approach by $k_c \sim \sqrt{g}T$ 
as discussed above, the factor $1/2$ in Eq.(\ref{eq:sstarhtl}) amounts
to the approximation $n_F(p) \simeq 1/2$ applied in the equilibrium 
calculation.
Performing the phase space integration for large $E_\gamma \gg T$ we find  the
soft contribution to the thermal photon rate
\begin{eqnarray}\label{eq:res.soft}
 \left. E_\gamma \frac{dR}{d^3 q} \right|_{soft} = e_q^2 \frac{3\alpha}{4\pi^3}
  \lambda_q m_q^2  e^{-E_\gamma/T}
  \ln \left[ \frac{k_c^2}{2m_q^2(\lambda_q,\lambda_g)} \right].
\end{eqnarray}

As in the equilibrium calculation the modified thermal mass provides 
a self-consistent IR cut-off under the logarithm. 
It is crucial to note the combination of fugacities present in 
Eq.(\ref{eq:res.soft}). With the appropriate changes in the thermal
mass according to Eq.(\ref{eq:mfne}) the same result is obtained
with J\"uttner distributions Eq.(\ref{eq:juettner}). For the case of
an equilibrium chemical potential $\mu < T\ll E_\gamma$ the 
multiplicative fugacity factor $\lambda_q$ in (\ref{eq:res.soft}) resulting
from the hard fermion line would be
absent \cite{thoma:muphoton}. The partial rate from the hard part of
phase space to which we turn next has to match this structure
in order that all singularities are  screened.

\subsection{hard part} \label{hard part}
For a discussion of the partial rate resulting from the region of hard
exchanged/loop momentum it is convenient to introduce Mandelstam 
invariants from the momentum variables 
defined in Fig.\ref{fig:fix} as
\begin{equation} \label{eq:man}
 s=(k+q)^2, \quad t=p^2 \qquad \mbox{for}\quad
  q + \bar q \to g+ \gamma,
\end{equation}
and accordingly for the Compton process. In terms of
these the kinematic region that remains to be considered is
expressed as
\begin{equation}\label{eq:phsphard}
   -s+k_c^2 \le p^2_{hard}=t \le -k_c^2  , \qquad 
  2k_c^2 \le s \le \infty .
\end{equation}
Here the calculation is to be
done in fixed order $e^2 g^2$, i.e., from the diagram 
shown in Fig.\ref{fig:fix}. We need the
quark propagation with one self-energy correction. 
For real photon production this amounts
to keeping  only the second term in  Eq.(\ref{eq:sres}).
The corresponding propagator is obtained 
by direct calculation of
\begin{equation} 
 i\delta S_{12}(p) = iS^0_{1a}(-i\Sigma_{ab}) iS^0_{b2},
\end{equation}
using the relations 
(\ref{eq:sigma}), or alternatively from an expansion of the resummed result
Eq.(\ref{eq:deltas.star}).  
At $O(g^2)$ the correction to the propagator is found to be 
\begin{eqnarray}
 \delta S_{12} (p) &=&
 -  \tilde n(p_0) (\Delta_R^2 -  \Delta_A^2) \slash p Re \Sigma(p) \slash p 
 - \frac{1}{2} \tilde n(p_0) (\Delta_R^2 + \Delta_A^2) 
 \slash p  (\Sigma_{12} - \Sigma_{21}) \slash p \nonumber\\ \label{eq:delta}
 & & - \Delta_R \Delta_A \slash p
  [ (1-\tilde n(p_0)) \Sigma_{12} + \tilde n(p_0) \Sigma_{21} ] \slash p.
\end{eqnarray}
Here the first term, which is  proportional to 
\begin{equation}
 (\Delta_R^2 - \Delta_A^2) = 2\pi i \varepsilon(p_0) \delta'(p^2) ,
\end{equation}
corresponds to the virtual correction to the quark propagator
and does not contribute to real photon production,
because the delta function has no support within the relevant region of 
phase space (\ref{eq:phsphard}). 

The second term would provide the result found in 
\cite{kapusta:photon,baier:photon}, when evaluated in equilibrium. Being
proportional to $Im \Sigma$ it corresponds to a vertical cut of the
diagram Fig.\ref{fig:fix}. 
Switching to Mandelstam variables $s$ and $t=p^2$ the factor 
\begin{equation}\label{eq:pp}
 \Delta_R^2 + \Delta_A^2 = 2{\bf P} \frac{1}{t^2} ,
\end{equation}
appearing here induces a pole in the t-channel of both
the Compton and annihilation process, which is to 
be regularized in the sense of a principal value. Together with
one extra factor of t from the spin structure of the quark propagators 
this term is at the origin of the logarithmic mass singularity cut-off
by $k_c^2$, 
\begin{equation} \label{eq:irsing}
 \int_{-s+k_c^2}^{-k_c^2} dt \frac{t}{t^2} = \ln
 \left(\frac{s-k_c^2}{k_c^2}\right). 
\end{equation}

The third term in Eq.(\ref{eq:delta}) contains the pinch singular 
contribution Eq.(\ref{eq:pinch}). 
It can be observed though that within the region of phase space under
consideration,  Eq.(\ref{eq:phsphard}), which does not include
the zeros of either $s,t,u$ the common pole position is 
approached from $t<0$ only. In this situation both 
pinch and principal value turn out to be  equivalent as
the regularization parameter $\varepsilon$ can be taken to zero:
\begin{equation}
\lim_{\varepsilon\to 0} \int_{-s+k_c^2}^{-k_c^2} dt \frac{t}{t^2+
   \varepsilon^2} = 
 {\bf P} \int_{-s+k_c^2}^{-k_c^2} dt \frac{t}{t^2} = 
  \int_{-s+k_c^2}^{-k_c^2} dt \frac{1}{t} .
\end{equation}
As a result  
associated terms containing the distribution $\tilde n(p_0)$ cancel each other.
The one-loop correction to the quark propagator  
therefore reduces to
\begin{equation}
 \left. \delta S_{12}(p) \right|_{p^2 \le -k_c^2} 
   \stackrel{\wedge}{=}
   -\frac{1}{(p^2)^2} \slash p \Sigma_{12} \slash p,
\end{equation}
which agrees with the corresponding expression in thermal equilibrium up 
to the fact that $\Sigma_{12}$ is now to be evaluated with the more 
general propagators for the present situation. Calculating
\begin{equation} \label{eq:pihard}
  -i\Pi_{12} (q) = - e^2 \int \frac{d^4p}{(2\pi)^4} 
  Tr\left[ \gamma_{\mu} iS_{12}(p-q)
   \gamma^\mu i\delta S_{21}(p) + S \leftrightarrow \delta S \right] ,
\end{equation}
we can again rely to a large extent on the work done before 
in the equilibrium calculation.

$\Pi_{12}$ of Eq.(\ref{eq:pihard}) contains contributions from 
both annihilation and Compton processes. 
The phase space 
integration for both partial rates may be written \cite{greiner:photon}
\begin{eqnarray} \label{eq:I1}
  I &=& \int_{2k_c^2}^\infty ds \int_{-s+k_c^2}^{-k_c^2} \frac{dt}{s}
  F(s,t) \Phi (s,t,E_\gamma; \lambda_q, \lambda_g),
\end{eqnarray}
where the matrix  elements $F(s,t)$ are the same as in equilibrium.
In $\Phi(s,t,E_\gamma;\lambda_q,\lambda_g)$ for the kinematic 
variables a number of integrations can be performed \cite{greiner:photon} 
resulting in boundaries $B_{1,2}$ for the remaining two dimensional 
integrations:  
\begin{equation} \label{eq:phi1}
 \frac{1}{s} \Phi(s,t,E_\gamma;\lambda_q,\lambda_g) = 
  \int_{B_1} dE_1 n_1 (E_1) \int_{B_2} dE_2 n_2(E_2) 
  (1\pm n_3(E_1+E_2 - E_\gamma)) (aE_2^2 + bE_2 +c)^{-1/2}
\end{equation}
with $a,b,c$ functions of $E_1$.
Here the actual combination of distribution functions, 
$n_i = n (n_i=\tilde n)$,
differs for the two processes and the sign in the enhancement (blocking) 
factor appears correspondingly. 
The expression Eq.(\ref{eq:phi1}) has been evaluated in previous work 
\cite{kapusta:photon,greiner:photon}
assuming Boltzmann approximation
for the incoming particles with energies $E_1, E_2$.

In order to obtain an IR finite emission rate it is 
crucial to keep full quantum statistics for the incoming 
particles as well in contrast to the treatment in 
\cite{thoma:nephoton,kaempfer:nephoton,muller:therma,greiner:muphoton,strickland:nephoton}. 
To demonstrate this we first concentrate
on the singular contribution alone which we isolate with the
help of a partial integration over $d\hat t = - dt/s$ in Eq.(\ref{eq:I1}).
The  contribution from the $t$-pole in the annihilation process 
can be written in the limit $k_c \to 0$: 
\begin{eqnarray}
 I_{t-pole}^{ann}&=&\int_0^\infty ds \left\{ \ln \left( \frac{s}{k_c^2} \right) 
  \Phi(s,t=0,E_\gamma; \lambda_i) - \Phi(s, t=-s, E_\gamma;\lambda_i) \right\} 
  \nonumber \\ \label{eq:I2}
 & &\qquad - \int_0^\infty ds \int_0^1 d\hat t \left[ \ln \hat t - \hat t
 \right] \frac{d}{d\hat t} \Phi(s,\hat t, E_\gamma;\lambda_i) . 
\end{eqnarray}
The logarithmic 
singularity present in the first contribution is seen to be 
related to processes with zero momentum transfer $t=0$ as expected. 
This restricts the energy of one incoming particle to coincide
with the energy of the unobserved final particle:
\begin{equation} \label{eq:phit}
 \Phi_{ann} (s,t=0,E_\gamma) = \frac{\pi^2}{E_\gamma} \tilde n(E_\gamma)
  \int_{s/4E_\gamma}^\infty dE \tilde n(E) (1+  n(E)). 
\end{equation}  
The same expression is found for the phase space $\Phi_{ann}(s,u=0,E_\gamma)$ 
weighting the singular
contribution from the $u$-pole in the annihilation process.
The Compton process on the other hand contributes two singular terms
from quark as well as anti-quark scattering each weighted by
\begin{equation} \label{eq:phiu}
 \Phi_{com}(s,t=0,E_\gamma) = \Phi_{com}(s,u=0,E_\gamma) = 
  \frac{\pi^2}{E_\gamma} \tilde n(E_\gamma) \int_{s/4E_\gamma}^\infty
  dE  n(E)  (1-\tilde n(E)).
\end{equation} 
Summing finally all four singular contributions, products of distribution
functions under the remaining integral cancel between the two processes
leaving behind the same linear combination $2[n(E)+\tilde n(E)]$ as it
appeared in the definition of 
the thermal mass Eq.(\ref{eq:mfne}). This is crucial for the latter to
provide the necessary cut-off. 
Treating quarks and anti-quarks differently for the case of an equilibrium
chemical potential the fermionic distributions in Eqs.(\ref{eq:phit}) and
(\ref{eq:phiu}) have to be adjusted correspondingly summing again
to $2n_B(E) + n_F(E-\mu) + n_F(E+\mu)$ defining the thermal mass
in this case. 
On the other hand from Eqs.(\ref{eq:phit}) and (\ref{eq:phiu}) it is also 
evident that applying the Boltzmann approximation to the incoming particles
spoils this nice feature in either case, and therefore the singularity
remains under this approximation. 

Interchanging the remaining two 
integrations over $s$ and $E_1$ and  again assuming the photon energy to 
be large with respect to  $T$, i.e., $\tilde n(E_\gamma) \simeq \lambda_q
e^{-E_\gamma/T}$, 
the singular part of the rate in the off-equilibrium situation 
is now easily obtained as
\begin{equation} \label{eq:ressing}
  \left. E\frac{dR^{sing}}{d^3 q} \right|_{hard} = 
    e_q^2  \frac{3\alpha}{4\pi^3} \lambda_q m_q^2 e^{-E_\gamma /T}  \ln 
    \left(\frac{4E_\gamma T}{k_c^2}\right) . 
\end{equation}

In order to complete the calculation we treat
the remaining two contributions in Eq.(\ref{eq:I2}) in order to 
fix the constant under the logarithm in Eq.(\ref{eq:ressing}).
This requires the unrestricted kinematic phase space 
$\Phi(s,t,E_\gamma;\lambda_q,\lambda_g)$ to be calculated with full 
quantum distributions. 
This can be accomplished by expanding Bose and Fermi distributions
as 
\begin{equation}
 n_F(p_0) = \sum_{n=1}^\infty (-)^{n+1} e^{-n \beta p_0 }, \qquad
 n_B(p_0) = \sum_{n=1}^\infty e^{-n \beta p_0 } .
\end{equation}
For the equilibrium case the resulting extra contributions
have been shown to cancel between the annihilation and Compton 
channel \cite{baier:photon} making the Boltzmann approximation for
the final result better than it is to be expected on the basis of 
individual processes \cite{kapusta:photon}. This cancellation,
however, depends on the properties of the equilibrium Fermi-Dirac
and Bose-Einstein distributions and is spoiled for the
off-equilibrium situation. For the specific choice Eqs.(\ref{eq:bdist}) and 
(\ref{eq:bdistf}) the additional contributions can nevertheless be 
evaluated without having to resort to classical approximations. 
We choose to order the resulting terms according to the powers
of fugacities present, having in mind that eventually  $\lambda_q \ll 
\lambda_g$. Taking out one common factor $\lambda_q$ we obtain 
for the remaining contribution
\begin{eqnarray} 
 \left. E_\gamma \frac{dR^{const}}{d^3q} \right|_{hard} &=&
    e_q^2  \frac{2 \alpha \alpha_s}{\pi^4} e^{-E_\gamma /T} T^2 \lambda_q 
  ~C(E_\gamma, T, \lambda_q, \lambda_g),
\end{eqnarray}
with
\begin{eqnarray}
  C(E_\gamma, T, \lambda_q, \lambda_g) &=&  \lambda_q 
  \left[ -1 + (1-\frac{\pi^2}{6}) \gamma + 
  (1-\frac{\pi^2}{12}) \ln\frac{E_\gamma}{T}  + \zeta_- \right]  
  \nonumber\\
  & & \quad + \lambda_g \left[ \frac{1}{2} + (1-\frac{\pi^2}{3})\gamma + 
   (1-\frac{\pi^2}{6} )\ln\frac{E_\gamma}{T} - \zeta_+ \right] 
  \nonumber \\ \label{eq:resconst}
  & &\quad + \lambda_q\lambda_g  \left[ \frac{1}{2} - \frac{\pi^2}{8} + 
   (\frac{\pi^2}{4} -2) (\gamma+\ln\frac{E_\gamma}{T}) + \frac{3}{2} \zeta'(2) 
  + \frac{\pi^2}{12}\ln 2 + (\zeta_+ - \zeta_-) \right] .  
\end{eqnarray}
The appearing symbols are Euler's constant $\gamma$, the derivative of the
Riemann $\zeta$-function evaluated at $\zeta'(2)$ and the sums
\begin{equation}
 \zeta_+ = \sum_{n=2}^\infty \frac{1}{n^2} \ln(n-1) \simeq 0.67, \qquad
 \zeta_- = \sum_{n=2}^\infty \frac{(-)^{n}}{n^2} \ln (n-1) \simeq -0.04.
\end{equation} 
The result Eq.(\ref{eq:resconst}) is successfully checked for consistency
by noting that the constants present in the equilibrium result 
\cite{baier:photon} are reproduced for $\lambda_q=\lambda_g=1$. 

Summarizing this subsection we find the partial rate for photon
emission from the hard part of phase space, Eq.(\ref{eq:phsphard}), to be
given by the sum of the two contributions Eq.(\ref{eq:ressing}) and
(\ref{eq:resconst}),
\begin{equation}\label{eq:reshard}
 \left. E\frac{dR}{d^3 q} \right|_{hard} = \left. E\frac{dR^{sing}}{d^3 q} 
  \right|_{hard} + \left. E\frac{dR^{const}}{d^3 q} \right|_{hard}.
\end{equation}

\section{result and conclusions} \label{result}

Calculating the complete emission rate for real photons as the
sum of the two partial rates Eqs.(\ref{eq:res.soft}) and (\ref{eq:reshard})
we see that the dependence on the arbitrarily chosen $k_c$ cancels. 
In the process the generalized thermal mass is 
established as self consistent cut-off for the logarithmic singularity
thereby justifying the approach taken in previous work.
Our result can be written
\begin{equation}\label{eq:resc}
  E_\gamma \frac{dR}{d^3q} = e_q^2 \frac{\alpha\alpha_s}{2\pi^2} \lambda_q T^2
   e^{-E_\gamma/T} \left[ \frac{2}{3}(\lambda_g + \frac{\lambda_q}{2})
  \ln \left(\frac{2E_\gamma T}{m_q^2} \right) + \frac{4}{\pi^2} C(E_\gamma ,
  T, \lambda_q, \lambda_g) \right] .
\end{equation}

The dynamical screening of the mass singularity actually does not depend on 
the explicit form, Eqs.(\ref{eq:bdist}) and (\ref{eq:bdistf}), of the 
distribution functions assumed in Eq.(\ref{eq:resc}).
In fact it only relies on the crucial combination of distributions
used to define the thermal mass in Eq.(\ref{eq:mfne}) to appear
in front of the singular factor from the sum of Eq.(\ref{eq:phit}) and
(\ref{eq:phiu}). In particular J\"uttner distributions 
Eq.(\ref{eq:juettner})  
lead to a result differing in the thermal mass parameter and
the constant factors under the logarithm, but the photon spectrum is 
equally free of singularities. 
The same conclusion holds also for an equilibrium situation with 
a non-vanishing chemical potential \cite{thoma:muphoton}. 

The second important result of our analysis is that we show explicitly 
the absence of
any additional contributions from pinch singular terms to 
the off-equilibrium photon production rate (\ref{eq:resc}). 
As far as the soft momentum scale contribution is concerned 
these terms are shown to be subleading with respect to the
dominant HTL-contributions. For the hard scale on the other hand
the absence of pinch singular contributions is due to 
the restricted kinematics of real photon production and does not
apply in general, e.g., for virtual photon production
the relevant region  of phase space is sufficiently enlarged for
these terms to become relevant \cite{baier:pirho,lebellac:pinch}.

Turning finally to more quantitative conclusions we note that the rate
Eq.(\ref{eq:resc}) differs in details from the results of both
Refs.\cite{shuryak:hotglue} and 
\cite{thoma:nephoton}; but 
indeed justifies the cut-off prescription employed therein. The
discrepancy with respect to 
\cite{thoma:nephoton} can be attributed to the fact 
that  distributions for the incoming particles 
have been treated in Boltzmann approximation. We find however that it is 
necessary to keep track of full quantum statistics in order for the screening
to be complete. 
Following \cite{shuryak:hotglue} we finally obtain a handy estimate
by restricting to the idealized situation of a fully saturated 
gluon distribution $\lambda_g=1$ with a quark admixture below its
equilibrium value by a factor $\lambda_q= \zeta = 1/4$
\footnote{ Under this condition the photon rate given in 
\cite{thoma:nephoton,kaempfer:nephoton,muller:therma,strickland:nephoton}
is dominated by the hard Compton process with the cut-off set by 
$k_c^2 = 2 m_q^2$, 
\begin{displaymath}
  E_\gamma \frac{dR}{d^3q} \simeq  e_q^2 \frac{2 \alpha\alpha_s}{\pi^4} T^2 
 e^{-E_\gamma/T} 
 \zeta \left[ \ln\frac{18 E_\gamma}{g^2 T} + \frac{1}{2} - \gamma \right]
  + O(\zeta^2) . 
\end{displaymath}
}.
In this situation the photon rate is approximated by 
\begin{equation}\label{eq:resap}
 E_\gamma \frac{dR}{d^3q} \simeq  e_q^2 \frac{\alpha\alpha_s}{3\pi^2} T^2 
 e^{-E_\gamma/T} 
 (\zeta+\frac{\zeta^2}{2})  \ln\left(\frac{c E_\gamma}{g^2 T} \right).
\end{equation}
There is an overall factor 2 with respect to \cite{shuryak:hotglue} and
a rather large constant $c\simeq 5$ under the logarithm 
depending on the actual value of $E_\gamma/T$. 

According to (\ref{eq:resap}) the photon emission rate
is reduced by a factor of about $3/8$ with respect to the equilibrium
case; this reduction is however easily 
compensated by a higher plasma temperature.

\vspace{0.5cm} 
\subsection*{Acknowledgments} 
\noindent 
One of us K.~R. acknowledges partial support
of the Gesellschaft f\"ur Schwer\-ionen\-for\-schung (GSI) and 
of the Committee of Research Development 
(KBN 2-P03B-09908). 
This work is  
supported in part by the EEC Programme "Human Capital and Mobility",
Network "Physics at High Energy Colliders", Contract CHRX-CT93-0357.
M.~D. is supported by Deutsche Forschungsgemeinschaft.
\newpage

\bibliographystyle{prsty}
\bibliography{photon,noneq,pinch,dilepton,ftft,therma,htl,books}

\end{document}